# GIS: (Geographic Information System)

## An application for socio-economical data collection for rural area

Mr.Nayak S.K.
Head, Dept. of Computer Science
Bahirji Smarak Mahavidyalaya,
Basmathnagar, Dist.Hingoli. (MS),
India

Dr.S.B.Thorat
Director,
Institute of Technology and Mgmt
Nanded, Dist.Nanded. (MS),
India

Dr.Kalyankar N.V.
Principal
Yeshwant Mahavidyalaya, Nanded
Nanded (MS)
India

***Abstract*—The country India follows the planning through planning commission. This is on the basis of information collected by traditional, tedious and manual method which is too slow to sustain.**

**Now we are in the age of 21 th century. We have seen in last few decades that the progress of information technology with leaps and bounds, which have completely changed the way of life in the developed nations. While internet has changed the established working practice and opened new vistas and provided a platform to connect, this gives the opportunity for collaborative work space that goes beyond the global boundary.**

**We are living in the global economy and India leading towards Liberalize Market Oriented Economy (LMOE). Considering this things, focusing on GIS, we proposed a system for collection of socio economic data and water resource management information of rural area via internet.**

## I. INTRODUCTION

### A. What is GIS?

1. Defination: A Geographic Information System (GIS) is a computer –assisted system for the acquisition, storage, analysis and display of geographic data. The terms:

i) Geographic- Indicates that data items are known in terms of geographic co-ordinates.

ii) Information- Implies that data in GIS is being organized and processed to yield useful knowledge.

iii) System – Implies that a GIS is made up from several interrelated and linked components with different functions. (Bonham-Carter-1994).

2. Geographic data: - This represents real world phenomenon in terms of:

a) Their shape and position (with reference to a known co-ordinate system)

b) Their characteristic attributes - e.g. soil type, cost, population etc.

c) Their special interrelationships (topology) with each other (describes their linking) [Burrough and McDonnell-1998]. This information is organized in layers/ themes.

## II. SYSTEMS OF GIS

Cartographic Display System, Map Digitizing System, Database Management System, Geographic Analysis System, Image Processing System, Statistical Analysis System and Decision Supports System. (See Fig.1).

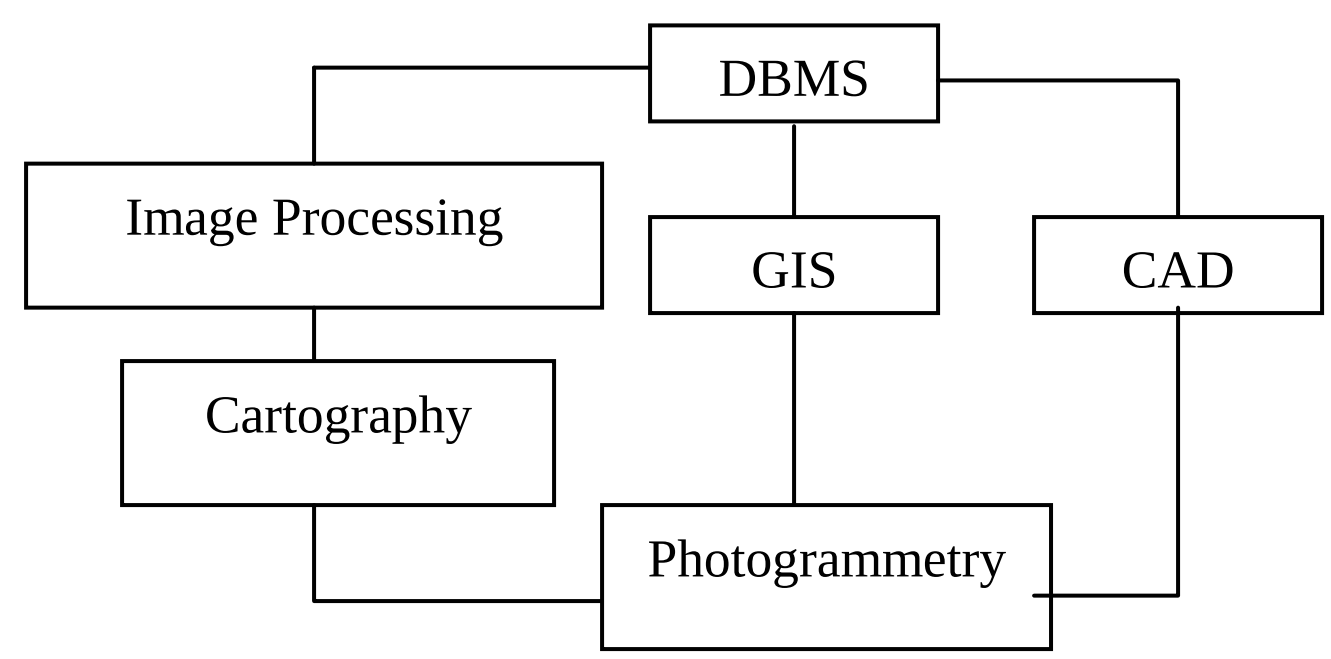

Figure 1. Systems of GIS

## III. COMPONENTS OF GIS

### A. computer hardware

Computer [Minimum P-4 1.5GHz based, [40 G.B. H.D., 128 MB RAM, CD Drive (52x), 1.44 MB drive, Color Monitor 14" Inbuilt-Modem, Sound Blaster and Ethernet Cards], Digitizer/Scanner, Plotter/Printers and Dialup connection through phone (i.e. internet hours) for inter-computer- access.

### B. Set of application software module

Application Software / Programme based on any GIS Software (e.g. ESR's- Arc View and Arc information, Intergraph (Geomedia), Auto-CAD Map, C-DAC's various software of GIS) as per need, Back-end and Front-end Software (e.g. Oracle and VB respectively), Web browser (e.g. Internet Explorer or Netscape Navigator). These modules need to be designed carefully after interaction with End-user to meet the requirements of application to be worth to be used by even a layman (i.e. it must be menu-driven and user friendly).

### C. Proper organizational contents(Burrough and McDonnell, 1998)

This refers to the management, analysis and other aspects involved in the implementation of GIS (The needs differ based on objective).

## IV. SOFTWARE

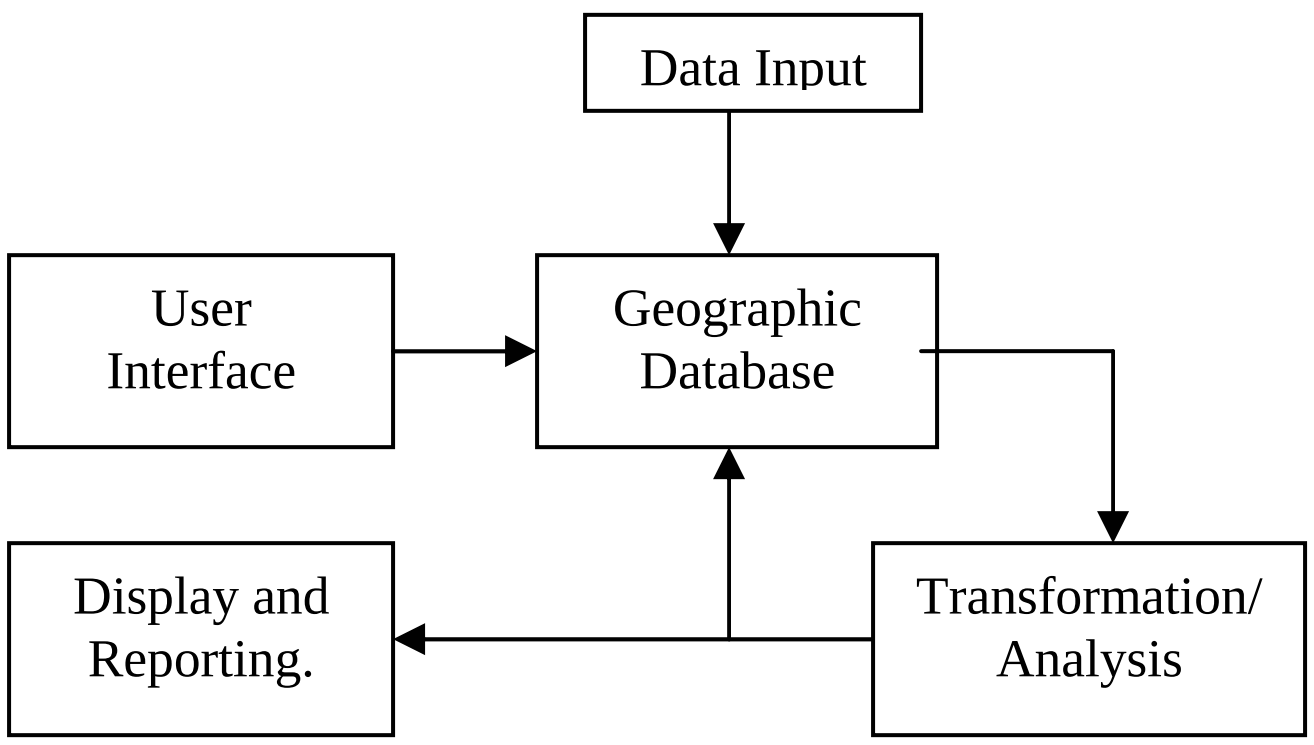

Figure 2. Main components of GIS(Burrough and McDonnell - 1998)

## V. SETTING UP A GIS

This includes integration of attribute data with graphical data. This also requires fieldwork, verification of the data and quality control. It is set up in four stages:

Stage 1: Planning- This is prerequisite and defines data requirment- Nature of data, its sources and its forms all defined attributes, graphic inputs such as maps or technical drawings, or external files from existing systems.

Stage 2: Preperation of Applications:- To prepare the applications and man-machine interfaces according characteristics defined in privious stage and with concent to end user.

Stage 3:- Data Capture:- Data is being collected from numerous sources, inputed to the system and verified. Begins at the conclusion of the planning stage and may be carried out with application stage. To meet the criteria defined in planning stage, quality control must be carried out.

Stage 4:- Implementation:- At this stage user gets acquented to accesss, to use and to manipulate the information, how to approch and how to solve the problems and to use Hardware and Software of GIS.

## VI. APPLICATION OF GIS

GIS has ever great impact on literally every field that can manages and analyses earth's surface related (i.e.Spatial) distributed data. Primarily aimed to integrate the scattered data and information with ever high speed and analyses it with accuracy in different departments.

Used in :-Agriculture, Archaeology, Environment, Health,Forestry, Navigation, Marketing, Real Estate, Regional and Local Planning, Road and Railway, Site Evaluation and Costing, Social Studies,Tourisim, Utilities etc.

## VII. OUR VIEW AND PROPOSED SYSTEM

### A. Our view

As we know India follows the planning through planning commission. Unfortunately this planning is on the basis of information collected by traditional, tedious and manual methods which are too slow to sustain and moreever when this so collected information reaches to the concern policy-maker, this information becomes absolute and so made policies could only yields nothing out of the most valuable efforts in terms of time, energy and money that is being invested. Hence, eventhough we are leading towards **Liberalize Market Oriented Economy (LMOE),** the importance of current, accurate, reliable and high speed data/information is very much identified and realized (e.g. industries like Hindustan Unlever Ltd, etc. are giving preference to this concept).

### B. Proposed idea

We are proposing herewith, the use of our own vast and reliable telephone network as a backbone (for dial up facility), exiting human resource with very basic training of computer ( must have been already trained for computer education by this time out of compulsion by Government ), the existing infrastructure of Grampanchayat , Panchayat Samities and Zilla-parishad ( some bit extra furniture and space could always be spared ) and will require the computer ( minimum P-4 1.5 based latest configuration ) at least one at each Grampanchayat and Panchayat Samiti or as per need.

An application software is required to be designed ( one time excersise ) and once being designed this could be issued to various Grampanchayats and Panchayat Samities as such. The concerned staff of Grampanchayat and Panchayat Samiti need only be able to read simple English text/identify icons (symbol) properly, since now a days most of the software are menu-driven and userfriendly ( hence most easy to be used even by lay man or novoice user as such).

Every Grampanchayat staff is being expected to have regerous surveys to collect the correct and reliable information of various fields/ attributes as per apllication and enter the same with time to time updations on concern computers which in turn are being connected via inbuilt modem and external telephone to the computers of other Grampanchayat and all Grampanchayats computers to concern Taluka's computer at Panchayat Samiti office so on, so that all Talukas computers are further connected to the concerned District (via dialup connection) . In this way merely seating at Taluka office, all necessary information that too accurately (**that too without going to concern villages**) can be collected, analysed and presented to head office (i.e.Z.P.) for further information and effective decesions as such.

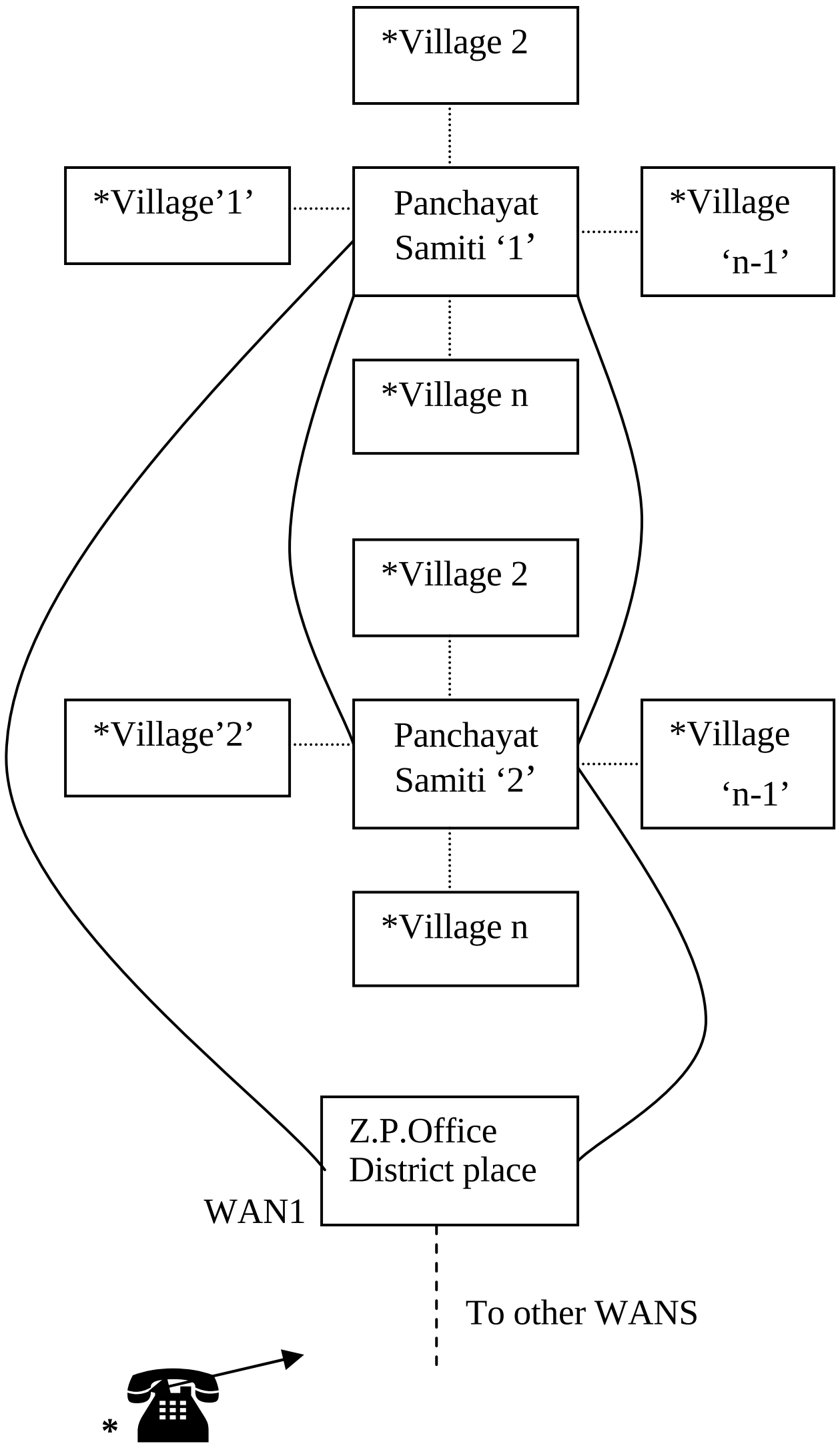

* Telephone and Computer at Concerned villages.

Figure 3. Proposed conceptual dial-up network for fast access of geographical data

Water is a vital resource for human survival and economic development; as populations and economies grow; water demand increases while the availability of the resource remains constant. Shortages engender water use conflicts, both in terms of quantity and quality. There is considerable variation across countries in laws and institutions related to water and planning and project implementation ability is not uniform. Therefore to have the uniformity in planning and project implementation ability, along-with socio-economic data, it is further proposed that the detail information of water resources, their current status, the contamination (if any), uses, capacity or level and other related issues could also be floated on internet for common sharing of Environment scientists for their analysis (attributes of this information could be selected in consultation with concern Environment Engineers and Scientists). In this way the current, accurate, reliable information regarding water resource management along- with socio-economic data can be made available for betterment and uplifts of rural area at speed beyond imagination.

The integrated management of other resources in the watershed, such as soil and vegetation, could also be included, as beneficiary participation in this data.

Efforts will be directed toward supporting water resources planning, policy making and management through required development proposed by concerned eminent scientists after analyzing this so made available data via internet. As per the intensities of the problem, this scheme will be an effective tool to take online decision as corrective measures whenever it will be required to do so.

The above proposed scheme is being needed to be deployed all over the country for volumenuous current and accurate geographical data collection, analysis and presentation of absolutely reliable inforamtion for further use.

Hence with the advent of technology the need of hour is to combine SCADA (Supervisory Control And Data Acquisition) with GIS for its wide sprade uses via internet for betterment of society as a whole, by giving the uniform, effective, useful, accurate and reliable geo-referenced/Socio-economic-data and water resource management information collection of rural area by completely automated system rather than by manual-data/information collection.

## VIII. MERITS AND DEMERITS

### A. Merits

This helps to have fast collection of socio-economic-data and water resource maangement information of rural area accurately and reliably. Hence could serve a strong basis for policy makers and certainly will lead to the betterment of society since it prevents sending wrong information from rural areas.

This model can be used for knowledge management, decision making tool and system for farmers.

If this system applied as online a huge amount of data can be done from portal that may helpful to farmers as well as government for future plannings and purposes.

If this system or model accepted and adopted by government and applied successfully, in some extent farmers suicide may be reduced.

### B. Demerits

Initial investment is high (but once being invested, recurring expenditure is only for the collection of data and analysing it, which could still otherwise would have been carried out manually).

## IX. CONCLUSION

Pertaining to need of hour, we must use this propesed scheme for collection of high speed, accurate and reliable data/information with ease to ensure proper planning to yield to have nation building policies. This potential system will certainly be useful and benifitial for betterment of rural areas and society as a whole.

## ACKNOWLEDGMENT

We are thankful to Hon. Ashok Chavan (Chief Minister, Maharashtra) India, Society members of Shri. Sharada Bhawan Education Society, Nanded. Also thankful to Shri. Jaiprakash Dandegaonkar (Ex-State Minister, Maharashtra), Society members of Bahiri Smarak Vidyalya Education Society, Wapti for encouraging our work and giving us support.

Also thankful to our family members and our students.

## AUTHORS PROFILE

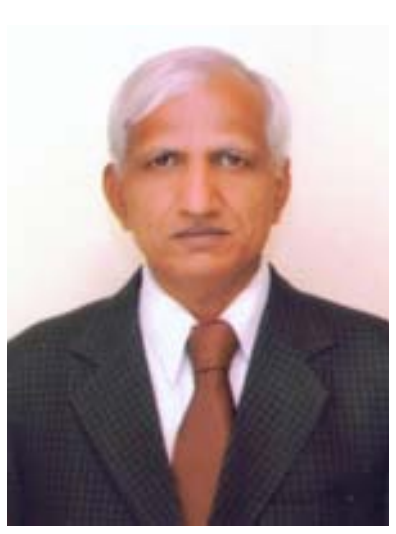

**Dr.N.V.Kalyankar**
Principal
Yeshwant Mahavidyalaya, Nanded (Maharashtra)

Completed M.Sc. (Physics) from Dr.B.A.M.U, Aurangabad. In 1980 he joined as a lecturer in department of physics at Yeshwant Mahavidyalaya, Nanded. In 1984 he completed his DHE. He completed his Ph.D. from Dr.B.A.M.U, Aurangabad in 1995. From 2003 he is working as a Principal to till date in Yeshwant Mahavidyalaya, Nanded. He is also research guide for Physics and Computer Science in S.R.T.M.U, Nanded. He is also worked on various bodies in S.R.T.M.U, Nanded. He also published research papers in various international / national journals. He is peer team member of NAAC (National Assessment and Accreditation Council, India). He published a book entitled “DBMS concepts and programming in FoxPro”. He also got “Best Principal” award from S.R.T.M.U, Nanded in 2009. He is life member of Indian National Congress, Kolkata (India). He is also honored with “Fellowship of Linnean Society of London (F.L.S.)” on 11 November 2009.

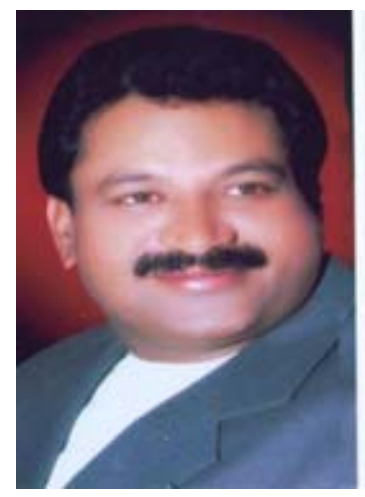

**Dr.S.B.THorat**
M.E. (Computer Science & Engg.)
M.Sc. (ECN), AMIE, LM-ISTE, Ph.D. (Comp.Sc. & Engg.)

He is having 24 years teaching experience. From 2001 he is working as a Director, at ITM. He is Dean of faculty of Computer studies at Swami Ramanand Teerth Marathwada University, Nanded (Maharashtra). Recently he is completed his Ph.D. He attended many national and International conferences. He is having 8 international publications. His interested area are AI, Neural network, Data mining, Fuzzy systems, Image processing.

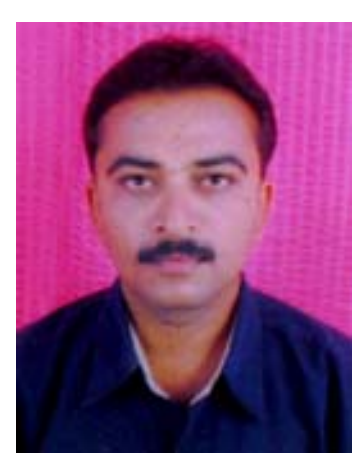

**S.K.Nayak**
M.Sc. (Computer Science), D.B.M, B.Ed.

He completed M.Sc. (Computer Science) from S.R.T.M.U, Nanded. In 2000 he joined as lecturer in Computer Science at Bahirji Smarak Mahavidyalaya, Basmathnagar. From 2002 he is acting as a Head of Computer Science department. He is doing Ph.D. He attended many national and international conferences, workshops and seminars. He is having 3 international publications. His interested areas are ICT, Rural development, Bioinformatics.